\documentclass[fleqn,10pt]{wlscirep}
\title{Variance reduction in output from generative AI}
\author[1,2]{Yu Xie}
\author[1,3]{Yueqi Xie}
\affil[1]{Paul and Marcia Center on Contemporary China, Princeton University, Princeton, NJ 08544, United States}
\affil[2]{Center for Social Research, Guanghua School of Management, Peking University, Beijing 100871, China}
\affil[3]{Hong Kong University of Science and Technology, Hong Kong}
\usepackage{color}
\usepackage{colortbl}
 \usepackage{CJKutf8}
\usepackage{bm}
\usepackage{hyperref}
\usepackage{xcolor}
\usepackage{lineno}
\usepackage{verbatim}
\usepackage{booktabs} 
\usepackage{algorithm}
\usepackage{algorithmic}
\usepackage{epsfig}
\usepackage{amsmath}
\usepackage{amsthm}

\usepackage{color}
\usepackage{amsmath}
\usepackage{graphicx,subfigure}
\usepackage{multirow}
\usepackage{makecell}
\usepackage{amssymb}
\usepackage[most]{tcolorbox} 
\tcbuselibrary{theorems} 

\usepackage{threeparttablex}
\newtcolorbox{system}[1]{
  colback=blue!5,
  colframe=blue!35!black,
  fonttitle=\bfseries,
  }
    \newtcolorbox{systemgp}[1]{
  colback=blue!5,
  colframe=blue!35!black,
  fonttitle=\bfseries,
  title={System-Mode Self-Reminder G.P. 1},
  }
  \newtcolorbox{systemop}[1]{
  colback=blue!5,
  colframe=blue!35!black,
  fonttitle=\bfseries,
  title={System-Mode Self-Reminder O.P.},
  }
\newtcolorbox{reasoner}[1]{
  colback=blue!5,
  colframe=blue!35!black,
  fonttitle=\bfseries,
  title={Reasoner},
  }
\newtcolorbox{refiner}[1]{
  colback=blue!5,
  colframe=blue!35!black,
  fonttitle=\bfseries,
  title={Refiner},
  }
  \newtcolorbox{prompt}[1]{
  colback=blue!5,
  colframe=blue!35!black,
  fonttitle=\bfseries,
  }
\newtcolorbox{response}[1]{
  colback=green!5,
  colframe=green!35!black,
  fonttitle=\bfseries,
  title={ChatGPT Defended by Self-Reminder},
  }

  \usepackage{subfigure}
\begin{abstract}

Generative AI models, such as ChatGPT, will increasingly replace humans in producing output for a variety of important tasks.  While much prior work has mostly focused on the improvement in the average performance of generative AI models relative to humans’ performance, much less attention has been paid to the significant reduction of variance in output produced by generative AI models. In this Perspective, we demonstrate that generative AI models are inherently prone to the phenomenon of “regression toward the mean” whereby variance in output tends to shrink relative to that in real-world distributions.  
We discuss potential social implications of this phenomenon across three levels—societal, group, and individual—and two dimensions—material and non-material.
Finally, we discuss interventions to mitigate negative effects, considering the roles of both service providers and users. Overall, this Perspective aims to raise awareness of the importance of output variance in generative AI and to foster collaborative efforts to meet the challenges posed by the reduction of variance in output generated by AI models.

\end{abstract}
\begin{document}
\begin{CJK*}{UTF8}{gbsn}
\flushbottom
\maketitle
%
%
\clearpage

Traditional artificial intelligence (AI) based on logical optimization algorithms will yield identical solutions given the same input conditions and is generally limited to certain well-defined tasks. The recent advent of generative AI models~\cite{touvron2023llama,link_chatgpt,chowdhery2022palm,openai2023gpt4,claudemodelcard}, learned from large bodies of observed data, has fundamentally advanced the field by generating AI output within any specific, concrete contexts and thus allowing for heterogeneous solutions. 
This is particularly well illustrated in the case of chatbots equipped with large language models (LLMs), such as ChatGPT~\cite{link_chatgpt}, which attracted more than 100 million users in the first two months after its launch.

Before generative AI models become widely accepted as standard tools for performing tasks important to human society, it is essential that we understand and carefully evaluate the output of such models. While previous work has mostly focused on the improvement in the average performance of generative AI models relative to human performance~\cite{klang2023evaluation,kung2023performance,liu2023llmrec,guo2023evaluating}, much less attention has been paid to the significant reduction in the variance of the output produced by such models. In this Perspective, we demonstrate that generative AI models are inherently prone to the phenomenon of “regression toward the mean,” whereby output variance tends to shrink relative to that in real-world distributions. We illustrate this problem with two simple empirical examples: the prediction of individual earnings in the US and the generation of abstracts for scientific papers.

We consider this tendency of variance reduction in generative AI output to be a logical result of the essential tension between the need to improve average prediction quality and the necessary simplification of real-life situations into a parameterizable space (albeit with an enormous number of parameters).

Following this, we explore the potential social implications of this phenomenon across three levels—societal, group, and individual—and two dimensions—material and non-material. Specifically, we emphasize four key aspects: (1) creativity and innovation, (2) stereotypes and statistical discrimination, (3) knowledge and information, and (4) individual identity. 
Finally, we explore potential interventions to mitigate these effects, considering the roles of both service providers and users.
In sum, our objective is to alert researchers to the tendency of variance reduction in the output of generative AI and how it may impact society in order to raise awareness and contribute to the development of potential solutions.

\section*{Two Examples Illustrating Variance Reduction}
We illustrate the phenomenon of variance reduction in the output of generative AI models with two examples. In both examples, we simulate real-life uses of generative AI models by providing prompts for certain tasks. In the first example, we ask ChatGPT to generate a numerical prediction of an American individual’s income based on varying levels of sociodemographic information. In the second example, we ask ChatGPT to generate an abstract from inputs with varying levels of information. Both examples clearly show the tendency of regression toward the mean, or alternatively, the reduction of variance, in generative AI output.

\begin{figure}[t]
    \centering

    \includegraphics[width=0.95\linewidth]{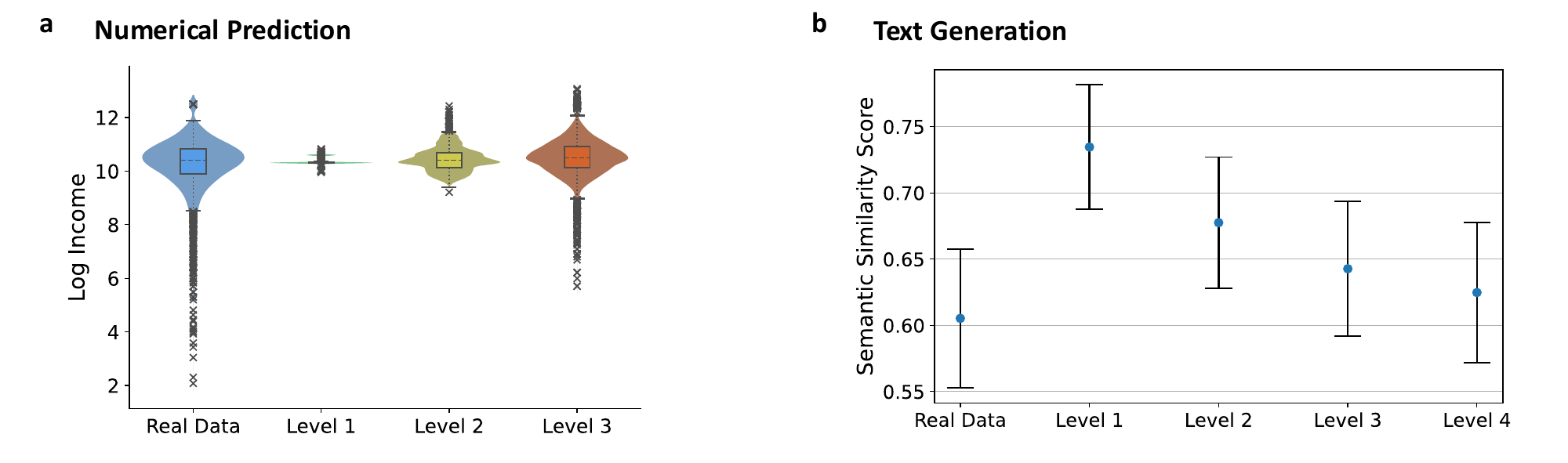}
    \caption{\textbf{a}. Distributions of the logarithm of 2004 income for real-world data and ChatGPT-predicted data, conditioned on different levels of information in the NYSL79 dataset. \textbf{b}. Means and standard deviations of semantic similarity scores for real-world data and ChatGPT-generated paper abstracts, conditioned on different levels of information from ArXiv.}
    \label{fig:overall}
\end{figure}

\subsection*{Example 1: Numerical Prediction}
We begin with numerical prediction of income, a key indicator of economic well-being in American society. We examine distributions of individual-level incomes predicted by ChatGPT under varying input conditions and compare them with the actual observed distribution.
For this exercise, we tasked ChatGPT with predicting person-specific 2004 income data for 6,041 individual respondents in the sample of the National Longitudinal Survey of Youth 1979 (NLSY79)~\cite{nlsy}, using varying levels of increasingly more detailed individual information: basic demographic information in Level 1, including age, race, and gender; occupation in addition in Level 2; and past income data from 1994, 1996, 1998, 2000, and 2002 in addition in Level 3.

We present the results in Figure~\ref{fig:overall}\textbf{a}. When only basic demographic information (Level 1) is used, the variance in the model’s output is significantly less than that of the actual data. This occurs because all sample units sharing the same basic demographic characteristics are given essentially the same mean-centered responses under these conditions.
Only when sufficient individual-level contextual information is provided, as in Levels 2 and 3, does the model begin to capture greater variance at the individual level. Although the output still regresses toward the mean even for Levels 2 and 3, the combination of means across various conditions becomes more reflective of real-world variance as the conditions become more detailed and person-specific.

In general, these numerical results confirm the presence of regression toward the mean and the reduction of variance in the output of generative AI models.

\subsection*{Example 2: Text Generation}
For the second test, we turn to text generation. Although textual data lack a direct representation of mean and variance, we can assess diversity by examining semantic similarity between texts. Specifically, for each text, we use Sentence-BERT~\cite{reimers2019sentence} to generate its semantic vector and compute the average cosine distance between that vector and the vectors of all other texts in the corpus. We then compute the similarity score by averaging the average cosine distance across all texts, with a higher similarity score indicating greater semantic similarity and thus lower diversity within the corpus.

We randomly sampled 2,000 scientific paper abstracts from the ArXiv dataset~\cite{clement2019use} and compared the diversity of the original data corpus with those of corpora regenerated by ChatGPT when varying the levels of input information. We used four levels of input information for abstract generation with ChatGPT: the subject of the article in Level 1, the title of the article in Level 2, a summary of the abstract in Level 3, and the full abstract in Level 4 with ChatGPT asked to refine it.

We present the results in Figure~\ref{fig:overall}\textbf{b}.
We observe that only when the full abstract is provided does the diversity of the generated content approach that of real-world data. In contrast, the generated corpora exhibit increasingly lower diversity (higher semantic similarity) at lower input levels, confirming the presence of regression toward the mean in text generation.

These findings suggest that when generative AI models such as LLMs are used for content creation without information-rich prompts, the output may lack the diversity and uncertainty typically associated with human generation.

\section*{Why Does Variance Reduction Occur in Generative AI Output?}

We consider variance reduction in generative AI output to be an inevitable result of\textit{ a fundamental paradox in generative AI}, with \textit{average accuracy being the main objective} of such models.

\subsection*{A Paradox in Generative AI}
The fundamental cause for the reduction of variance in generative AI output lies in the paradox between the infinite heterogeneity of human phenomena and the finite (albeit extremely large) number of parameters used to represent them in generative AI models. 

There are good reasons to believe in irreducible variability, i.e., infinite heterogeneity, at the individual level for almost all human phenomena~\cite{xie2013population}. Such variability may originate from either nature -- gene-based biological differences that give rise to evolution~\cite{mayr1982growth,darwin1859origin} or nurture -- social situations~\cite{galton1889natural} where individuals face different environments, such as the family, neighborhood, social network, and have different accumulative life-courses experiences.

In contrast, generative AI models do have a capacity bound by design. Their goal is to derive generalizable patterns that depend on a finite (albeit very large) number of parameters to represent real-world distributions as effectively as possible for any given input. 
To achieve this, AI models have to assume some level of homogeneity, simplify the distribution in the real world at the aggregate level characterized by a parameterized space
rather than at the individual level, and make inferences by drawing on ``similar situations'' in this extremely large, multi-dimensional space. 
This process inevitably overlooks the residual variability at the individual level not captured by the parameterized space, causing the model output to regress toward average patterns with reduced variance.

This necessary simplification limits the upper bound of the variance that AI models can represent, making it significantly lower than the variance that exists in real-life situations.

\subsection*{Average Accuracy as the Main Objective}
In addition to the inherent limitations posed by the capacity of generative AI models in representing real-life variance, commonly employed accuracy-oriented decoding strategies further amplify the phenomenon of variance reduction.

Consider the process of language modeling with AI model $M_\theta$, where the task is to recursively predict the next token until the end of the sequence.
Given the preceding context $x$, for each possible next token  $y_i$  in the vocabulary, the model computes an unnormalized score  $z_i$. These scores are then converted into probabilities using the following softmax function:
\begin{equation}
p_{\boldsymbol{\theta}}(y_i \mid x) = \frac{\exp(z_i)}{\sum_{j} \exp(z_j)}.
\end{equation}
During training,  $p_{\boldsymbol{\theta}}(Y|x)$ is trained to approximate an observed distribution $p(Y\mid x)$ derived from the training data.

During decoding (inference), the model needs to sample one token for the exact predicted next token. Instead of sampling from this learned $p_{\boldsymbol{\theta}}(Y \mid x)$, a common practice is to apply temperature ($\tau$) scaling~\cite{ackley1985learning} to adjust the sharpness of the next-token probability distribution: 
\begin{equation}
p^{\prime}_{\boldsymbol{\theta}}(y_i \mid x) = \frac{\exp\left(\dfrac{z_i}{\tau}\right)}{\sum_{j} \exp\left(\dfrac{z_j}{\tau}\right)}.
\end{equation}

In practice, to achieve a more stable average accuracy, the temperature  $\tau$  is typically set between $0$ and $1$. 
For example, the default temperature for models like LLaMA~\cite{touvron2023llama, touvron2023llama2} is often configured to values below 1.
A lower temperature reduces randomness in the output by concentrating the probability mass on the most likely tokens. In the extreme case, setting $\tau = 0$  corresponds to greedy decoding, where the model always selects the token with the highest probability.

Building on this, many models further incorporate additional techniques such as top-$p$  sampling~\cite{holtzman2019curious} and top-$ k $ sampling~\cite{fan2018hierarchical}.
These methods restrict the sampling process to the most probable tokens based on a cumulative probability threshold ($p$) or a fixed number of top tokens ($k$), ensuring that the model generates frequently observed results.

All of these decoding strategies capitalize on
high probabilities for generated results, thereby amplifying the variance reduction in AI-generative outputs. 
Although these approaches improve the stability of outputs, they come at the cost of sacrificing the diversity that AI models are capable of expressing.

\section*{Recovery of Variance with Fine-grained Contexts}
In the previous section, we discussed how variance reduction is an inevitable outcome of the capacity limitations of AI models and their primary focus on accuracy. 
In this section, we demonstrate how refining user input contexts can increase between-context variance, thereby partially recovering the overall variance.

To directly analyze numerical variance under different levels of context granularity, we consider AI output $Y$  to be a numerical prediction outcome.  Nevertheless, the concept can be generalized to other types of AI generation.
Suppose an input context  $c$.  We compare two possibilities, the original context $c=c_0$ and a further differentiated context  $c \in C = \{c_1, c_2, \dots, c_n\}$, which collectively represent possible specific versions of  $c$. We then examine how this refinement impacts the variance of output $Y$.

For the original context  $c_0$, the variance in output  $Y$  is formulated as  $\text{Var}(Y \mid  c_0)$. When the context  $ c$  is refined into a set of fine-grained contexts $ c \in C $, the total variance of  $Y$  conditioned on these refined contexts can be decomposed as follows:
\[
\text{Var}(Y \mid c \in C) = \text{Var}\left(\mathbb{E}[Y \mid c \in C]\right) + \mathbb{E}\left[\text{Var}(Y \mid c \in C)\right],
\]
where  $\text{Var}(\mathbb{E}[Y \mid c \in C])$  represents the \textit{between-context variance}, and $ \mathbb{E}[\text{Var}(Y \mid c \in C)] $ represents the \textit{within-context variance}.
The between-context variance measures the variation in the expected values of  $Y $ across different specific contexts  $c_1, c_2, \dots, c_n $, while the within-context variance measures the average variability of  $Y$  within each fine-grained context  $c_i$  in $C$.

Refining  $c$  into  $C$  introduces heterogeneity across the specific contexts, as the expected values  $\mathbb{E}[Y \mid c=c_1], \mathbb{E}[Y \mid c = c_2], \dots ,\mathbb{E}[Y \mid c = c_n]$ are likely to differ, contributing to a typically large between-context variance term,  $\text{Var}(\mathbb{E}[Y \mid c \in C])$.
At the same time, fine-grained contexts typically result in a small increase in within-context homogeneity, leading to a reduction in the expected within-context variance,  $\mathbb{E}[\text{Var}(Y \mid c \in C)] $, relative to  $\text{Var}(Y \mid c = c_0) $. 
However, this reduction is typically small and insufficient to counterbalance the increase introduced by the between-context variance. As a result, the total variance tends to increase with context refinement.  This explains our empirical findings that providing AI models with more fine-grained contexts leads to outputs with higher variance.

 

\section*{Social Implications}
We acknowledge that the social impact of reduced variance encompasses a wide range of issues, which cannot be exhaustively discussed in this Perspective. As an illustration, we focus on several selected aspects that we consider particularly important. Of course, future research can extend our limited discussion to other topics.  

We organize our discussion of the social implications of reduced variance in output generated by generative AI according to three levels—societal, group, and individual—and two dimensions—material and non-material.  The three levels correspond to distinct aspects of variance reduction: (1) the societal level: overall variance reduction in total distribution; (2) the group level: the compression of individual characteristics into group-level averages; and (3) the individual level: the loss of individual uniqueness due to an individual's characterization through parameterization. 
The material dimension emphasizes tangible outcomes like creative output, innovation, and decision-making processes, whereas the non-material dimension primarily focuses on aspects such as opinions, values, preferences, and other intangible human attributes.  

An overview of the discussion is provided in Table~\ref{tab:category}.




\begin{table}[t]
    \centering
    \begin{tabular}{l|ll}
    \toprule
    \textbf{}  &\textbf{Non-material}& \textbf{Material} \\
    \midrule
        \multirow{1}{*}{\textbf{Societal Level}}& Collective Knowledge and Information&  Collective Creativity and Innovation \\
                \midrule     
                        \multirow{1}{*}{\textbf{Group Level}}&Stereotype& Statistical Discrimination\\
        \midrule  
        \multirow{1}{*}{\textbf{Individual Level}}&Individual Identity & Individual Creativity and Innovation\\
    \bottomrule
    \end{tabular}
    \caption{Social implications of reduced output variance in generative AI, categorized across three levels (societal, group, and individual) and two dimensions (non-material and material).}
    \label{tab:category}
\end{table}

\subsection*{Societal Level}
The reduction in output variance of generative AI can lead to \textit{a decline in overall diversity} across various domains for a society as a whole.  The potential 
negative impacts can be either non-material, involving collective knowledge and information, or material, involving collective creativity and innovation.

\textbf{Collective Knowledge and Information:} 
Collective knowledge and information refer to the shared repository of facts, concepts, perspectives, and cultural practices maintained within a human society. As generative AI becomes an increasingly mainstream tool for knowledge and information acquisition~\cite{baidoo2023education,lo2023impact,IBTimesBing2024}, its inherent tendency toward variance reduction raises the possibility of narrowing this collective knowledge resource.
Compared to traditional methods of information acquisition, generative AI increases the likelihood that individuals will encounter similar, homogenized narratives. This can discourage the exploration of less common or minority perspectives, limiting exposure to diverse viewpoints.
As highlighted in prior work~\cite{peterson2024ai}, this trend risks suppressing long-tail information and progressively curtailing the richness of our shared knowledge compared to the broader breadth of historical knowledge.


\textbf{Collective Creativity and Innovation:} Variance is essential to fostering collective creativity and driving innovation in society. Research on \emph{functional diversity} suggests that diverse groups often outperform teams composed solely of the highest-performing individuals~\cite{hong2004groups,burton2024large,page2008difference,hong2012some}. This is because high-performing individuals often have similar characteristics, which can limit the range of perspectives and solutions within the group.
Generative AI models, while capable of producing highly accurate and effective ideas according to past experiences and accepted standards due to their training on vast datasets, generate outputs that are relatively homogeneous. 
As a result, individuals who rely heavily on these models tend to think ``in a box" rather than ``out of the box," and their outputs are at risk of having a high degree of similarity to those of other ``high-performing” contributors who are also relying on generative AI.
When large numbers of people depend on the same AI models for creative tasks—such as idea generation, refinement, or decision making—there is a risk that, despite the model’s high performance, this uniformity could reduce functional diversity among individuals, potentially stifling collective creativity and innovation at a societal level.



\subsection*{Group Level}
Second, the reduction in the output variance of generative AI leads to \textit{the compression of individual characteristics into group-level averages}, which can further result in the reinforcement of group-level stereotypes, a non-material implication, and statistical discrimination, a material implication.


\textbf{Stereotype:} 
A stereotype can be viewed as an oversimplified, or compressed average, representation of the (assumed) distribution of the characteristics of a group of people~\cite{kanahara2006review}. 
Stereotypes can be associated with race/ethnicity, gender, occupations, and various demographic categories, potentially negatively impacting individuals in socially disadvantaged groups~\cite{spencer2016stereotype,fiske2018model,steele1995stereotype}. 
The variance reduction inherent in generative AI, which tends to suppress within-group variability and gravitate toward group-level average representations, can further reinforce existing stereotypes~\cite{zhou2024bias, bianchi2023easily}. 
In language-based applications, LLMs are widely used for information retrieval and content generation~\cite{zhao2023survey,xie2024measuring,zhuang2023toolqa}, yet their outputs can embed and perpetuate stereotypes when describing individuals~\cite{ferrara2023should,kotek2023gender,abid2021persistent,gallegos2024bias,liubias}. For instance, gender biases in LLM-generated reference letters have been documented~\cite{wan2023kelly}, with phrases like “Kelly is a warm person” and “Joseph is a role model” reflecting stereotypical gender associations.
In multimodal applications, generative tools for audio, image, and video can “bring to life” content from textual descriptions~\cite{croitoru2023diffusion,goodfellow2020generative}, but they also risk perpetuating stereotypes. For instance, research has shown that widely used text-to-image models often amplify demographic stereotypes: for example, prompts like “Ethiopian man with his house” consistently generate depictions focused on poverty, with little variation~\cite{bianchi2023easily}.
As generative AI models become increasingly prevalent across these tasks, the risk of reinforcing group-level  stereotypes and biases continues to grow, potentially exacerbating existing inequalities.

\textbf{Statistical Discrimination:} Statistical discrimination is a material implication of the group-level stereotyping discussed above: for example, when such stereotyping informs the actions of employers. Consequential decision making based on group characteristics such as race, gender, or age, rather than individual attributes, can potentially result in systemic unfair treatment.
Statistical discrimination can be illegal under certain circumstances (e.g., gender-based labor market discrimination)~\cite{dickinson2009statistical}.
When generative AI is applied to decision making, variance reduction impairs the user's ability to evaluate individuals based on their unique characteristics, increasing the risk of defaulting to group-level attributes~\cite{hacker2024generative}. For example, in employment or lending decisions, generative AI may rely on summary statistics that reflect and amplify group-level differences across observed attributes in the training data. Such practices could potentially perpetuate existing inequalities and unfairly disadvantage certain demographic groups.

\subsection*{Individual Level}
Third, the reduction in the output variance of generative AI can lead to \textit{indifference to individuals} due to the loss of individual residuals, affecting both individual identity, a non-material implication, and individual creativity and innovation, a material implication.  

\textbf{Individual Identity:} 
Individual identity refers to a person's subjective self-assessment that  distinguishes them from others, including personal characteristics, preferences, social categories, attitudes, and values. When generative AI reduces variance by compressing representations into group averages, individual residuals—traits that set individuals apart but which are either not parameterized in or not input as prompts into AI models—are essentially discarded. While the residual variance in AI-generated outputs is random and algorithmic, real human residuals reflect fixed, meaningful attributes that may carry deep personal and social significance. For instance, religious beliefs, political views, and personal preferences are inherently individualistic and may not be fully captured or represented by AI models, despite being very important to people in real life. 
This systematic indifference to person-specific opinions, preferences, and values in generative AI outputs can subtly erode individual identity over time. As individuals interact with AI-generated content, their beliefs and opinions may be influenced, shaped, or even homogenized, resulting in a gradual loss of
individuality. Alternatively, some people may find AI-generated content unsatisfactory and stop using AI for content generation as a result of this. 



\textbf{Individual Creativity and Innovation:}
The impact of variance reduction in generative AI on individual creativity can be viewed from both positive and negative perspectives. On the positive side, generative AI, when applied appropriately, has the potential to enhance individual creative output. Studies have shown that generative AI can benefit individual creativity in specific contexts~\cite{si2024can,gao2024quantifying,lee2024empirical}. We know, for example, that it can help people avoid making both factual and stylistic errors.  
On the negative side, however, the ease of obtaining high-quality results can discourage individuals from striving for extraordinary breakthroughs. Historically, major cultural, intellectual, artistic, or scientific advancements have often been driven by the ``out of box" ideas of extraordinary individuals~\cite{cole1972ortega,sawyer2024explaining}. While generative AI serves as a convenient tool that directs users toward high-quality (yet standardized) results, there is a risk that individuals may lose the motivation for independent thought and original creativity. As a result, the potential for groundbreaking innovations could diminish, leaving society with an abundance of high-quality yet routine creations.

\begin{table}[t]
    \centering
    \begin{tabular}{l|ll}
    \toprule
    \textbf{}  &\textbf{Variance Preservation}& \textbf{Caution in Use} \\
    \midrule
        \multirow{2}{*}{\textbf{Service Provider}} & Comprehensiveness of Training Data&  Education and Warning\\
        &Model Personalization\\
                \midrule     
                        \multirow{1}{*}{\textbf{User}} &Specificity of Input & Reducing Over-reliance\\
    \bottomrule
    \end{tabular}
    \caption{Mitigation strategies for meeting the challenges associated with variance reduction in generative AI, categorized by roles (service providers and users) and their focus (variance preservation and caution in use).}
    \label{tab:mitigation}
\end{table}

\section*{Mitigation}
Having examined the potential social implications of variance reduction in generative AI, we propose a set of mitigation strategies—centered on \textit{variance preservation} and \textit{caution in use}—as summarized in Table~\ref{tab:mitigation}.
Although there is no universal solution for countering the tendency of variance reduction in generative AI output, we aim to provide actionable insights for \textit{service providers} and \textit{users}.


\subsection*{Service Provider}
\textbf{Comprehensiveness of Training Data:}
Service providers are encouraged to ensure that training data used in generative AI models are comprehensive and representative of diverse demographics, cultures, knowledge domains, and perspectives. This diversity in training data is crucial because generative AI models can only produce relevant and accurate content when they are sufficiently trained on data that encompass the corresponding knowledge and perspectives. Without such coverage, models struggle to reflect underrepresented information, regardless of how precisely users craft their inputs.
For example, research has shown that the performance of LLMs in processing knowledge from low-resource languages often falls behind their performance in English, primarily due to significant gaps in the availability of training data~\cite{jin2024better,huang2023improving}. Similarly, state-of-the-art LLMs struggle with local cultural knowledge, such as understanding Basque culture~\cite{etxaniz2024bertaqa}.
Therefore, assembling comprehensive training datasets serves as a critical foundation for enabling generative AI models to preserve variance in their outputs.

\textbf{Model Personalization:} 
Beyond efforts to enhance the representativeness of foundation models, enabling model personalization represents another approach to preserving variance in AI outputs. 
Typically, generative AI models are aligned with “common” human values and preferences, possibly excluding or under-representing certain groups or individuals~\cite{kirk2024benefits}. Personalization offers a potential solution, allowing users to adapt and shape AI models to suit their unique needs and preferences rather than relying on a one-size-fits-all approach. 
For instance, an LLM could be personalized to align with a user’s political inclinations, religious beliefs, and perspectives on various topics, thereby preserving their distinctive identities and contributing to essential variance at the societal level.
LLM personalization has been both theoretically explored in research~\cite{kirk2024benefits} and practically attempted by chatbot services like ChatGPT, which utilize users’ interaction histories to deliver more personalized responses.

\textbf{Education and Warning:} 
Beyond variance preservation, service providers can also mitigate negative effects through warnings and user education. Existing warnings in generative AI often focus on accuracy issues. For example, ChatGPT displays messages like “ChatGPT may make mistakes. Please verify important information,” alerting users to potential factual inaccuracies in AI output.
We suggest that service providers expand their warnings and educational efforts to also address variance reduction. For instance, they could inform users about the potential for output homogenization (variance reduction), whereby generated outputs may become similar across different users for the same prompted task. Additionally, service providers could educate users on how to craft more specific and tailored prompts to generate outputs that better align with their unique needs and perspectives. This approach would enhance both user awareness and engagement, helping to mitigate the negative effects associated with variance reduction.

\subsection*{User}
\textbf{Specificity of Input:}
User attention to the specificity of inputs plays a crucial role in preserving essential variance in AI output. 
As demonstrated in our empirical experiments and theoretical analysis, when input contexts are sufficiently fine-grained and rich in information, the outputs of generative AI can more closely reflect real-world variance.
In other words, while simple or generic inputs tend to produce more homogenized responses, detailed and nuanced prompts can unlock the full diversity potential of generative AI models by introducing heterogeneity as between-context variance.
Greater awareness of the need for such specificity at the individual level can not only result in more accurate and relevant answers, but at the societal level it can also help mitigate the reduction of variance in information, perspectives, and creative output.

\textbf{Reducing Over-reliance:}
Reducing over-reliance on generative AI is another proactive step users can take. While generative AI serves as a powerful and convenient tool for various applications, excessive dependence on it can lead to anchoring bias~\cite{tversky1974judgment,choi2024llmeffecthumanstruly}, where individuals rely too heavily on the initial output they receive from AI. 
Over-reliance on AI substitutes for humans' effort for further exploration and ultimately yields the negative impact of variance reduction discussed earlier.
To address this, users are encouraged to engage in independent exploration and critical thinking, ensuring that AI-generated content does not become the sole or dominant source of information or ideas.



\section*{Conclusion}
The average performance of generative AI has garnered significant research attention and its implications for various aspects of human society have been extensively analyzed. 
In this Perspective, we move beyond average performance to focus on the often overlooked variance of AI output. 
We highlight the phenomenon of regression toward the mean in generative AI, wherein output variance tends to shrink compared to real-world distributions.
We analyze this phenomenon as a natural consequence of the constraints imposed by AI model capacity and the prediction accuracy objectives of AI applications. 
We delve into the social implications of variance reduction at the societal, group, and individual levels, as well as across non-material and material dimensions.
Finally, we propose potential mitigation strategies to meet the challenges posed by variance reduction in AI output, offering actionable insights for both service providers and users.
We aim for this Perspective to draw research attention to the inherent variance reduction in generative AI and its broader social implications, thereby fostering collaborative efforts to minimize its potential negative impacts.
\bibliography{cost}






\end{CJK*}
\end{document}